# Title: Observation of a Nematic Quantum Hall Liquid on the Surface of Bismuth

**Authors:** Benjamin E. Feldman[1†], Mallika T. Randeria[1†], András Gyenis[1†], Fengcheng Wu[2], Huiwen Ji[3], R. J. Cava[3], Allan H. MacDonald[2], and Ali Yazdani[1]*

**Affiliations:**
[1]Joseph Henry Laboratories & Department of Physics, Princeton University, Princeton, NJ 08544, USA
[2]Department of Physics, The University of Texas at Austin, Austin, TX 78712, USA
[3]Department of Chemistry, Princeton University, Princeton, NJ 08544, USA

[†]These authors contributed equally to this work.
*Correspondence to: yazdani@princeton.edu

**Abstract:** Nematic quantum fluids with wavefunctions that break the underlying crystalline symmetry can form in interacting electronic systems. We examine the quantum Hall states that arise in high magnetic fields from anisotropic hole pockets on the Bi(111) surface. Spectroscopy performed with a scanning tunneling microscope shows that a combination of single-particle effects and many-body Coulomb interactions lift the six-fold Landau level (LL) degeneracy to form three valley-polarized quantum Hall states. We image the resulting anisotropic LL wavefunctions and show that they have a different orientation for each broken-symmetry state. The wavefunctions correspond to those expected from pairs of hole valleys and provide a direct spatial signature of a nematic electronic phase.

**Main text:**

Nematic electronic states represent an intriguing class of broken-symmetry phases that can spontaneously form as a result of electronic correlations (*1, 2*). They are characterized by reduced rotational symmetry relative to the underlying crystal lattice and have attracted considerable interest in systems such as two-dimensional electron gases (2DEGs) (*3-5*), strontium ruthenate (*6*), and high-temperature superconductors (*7-12*). The sensitivity of electronic nematic phases to disorder results in short range ordering and domains, making them difficult to study using global measurements that average over microscopic configurations. The effect of perturbations, such as crystalline strain, may be used to show a propensity for nematic order, *i.e.* to provide evidence that vestiges of nematic behavior survive even in the presence of material imperfections (*1*). However, it is difficult to quantitatively correlate the experimental evidence of ordering with a microscopic description of the electronic states and the interactions responsible for nematic behavior. To put the study of nematic electronic phases on more quantitative ground, it is therefore important not only to perform local measurements, but also to find a material system for which theory can fully characterize the underlying broken-symmetry states and the electronic interactions.

Multi-valley 2DEGs with anisotropic band structure have been anticipated as a model platform to explore nematic order in the quantum Hall regime (*13-17*). The key idea is that Coulomb interactions can spontaneously lift the valley degeneracy in materials with low disorder and thereby break rotational symmetry. In contrast to previously studied metallic nematic phases, this leads to a gapped nematic state with quantized Hall conductance. We examine such a 2DEG on the surface of single crystals of bismuth (Bi), which is one of the cleanest electronic systems, with a bulk mean free path reaching 1 mm at low temperatures (*18*). Interest in Bi has recently been rekindled by bulk measurements showing phase transitions and anisotropic behavior at large magnetic fields, which may be related to nematic electronic phenomena (*19-22*). We focus here on the (111) surface of Bi, for which strong Rashba spin-orbit coupling results in a rich 2DEG consisting of spin-split surface states that produce multiple electron and hole pockets (*23, 24*). Scanning tunneling microscope (STM) images (Fig. 1A) show that the in situ cleaved Bi(111) surface has large (> 200 nm *x* 200 nm) atomically ordered terraces that are separated by steps oriented along high-symmetry crystallographic directions (*25*). Angle-resolved photoemission spectroscopy (ARPES) measurements (*23, 24, 26, 27*) of this surface show that its Fermi surface consists of a hexagonal electron pocket at the Γ point, three additional elongated electron pockets around the M points, and six anisotropic hole pockets along the Γ-M directions (Fig. 1B, inset). The multiply degenerate anisotropic valleys and the low disorder of the Bi(111) surface make it an ideal system to search for nematic electronic behavior using the STM.

In the absence of magnetic field, spectroscopic measurements of the Bi(111) surface with the STM (Fig. 1B) show features in the tunneling conductance $G$ that are related to van Hove singularities of the density of states (DOS), such as the sharp peak at energy $E = 220$ meV and the abrupt drop at 33 meV. These features correspond to the upper band edges of the surface states along the Γ-M direction (*25, 28, 29*). In the presence of a large magnetic field $B$, the electron and hole states of the Bi(111) surface are quantized into Landau levels (LLs), each with degeneracy $geB/h$, where $e$ is the electron charge, $h$ is Planck's constant, and $g$ accounts for the degeneracy arising from the valley degree of freedom ($g = 6$ for holes). At high magnetic field, the STM spectra show a series of sharp peaks (Fig. 1B) whose evolution with magnetic field can be used to distinguish between electron- and hole-like LLs, which disperse in energy with positive or negative slopes, respectively, as a function of magnetic field (Figs. 1C and 1D). They do not exhibit avoided crossings, and the total conductance is additive when they cross, which suggests independent tunneling into each LL. LL spectroscopy on thin Bi(111) films was recently reported (*28*), but did not show evidence of symmetry breaking, which is the focus of our work.

Our first key observation is that the surface state LLs do not disperse linearly with magnetic field. Instead, they are pinned to the Fermi level until they are fully occupied, as is clearly shown for the hole states in Fig. 1D. Such behavior is rarely observed in LL spectroscopy of ungated samples performed using a STM (*30*), and it indicates that the surface charge density is held constant in our system. Electron LLs exhibit pinning only when there are no proximal hole states, whereas they otherwise cross straight through the hole LLs at the Fermi level. This difference in behavior signals an intriguing competition between electron- and hole-like states in a magnetic field, and suggests charge rearrangement between pockets (*29*). We focus below on the hole states, for which the orbital index $N_h$ is straightforward to assign, with the highest-energy peak corresponding to $N_h = 0$ closely matched to the zero-field drop in conductance at 33 meV. Using the values of the field and filling factor at which LLs cross the Fermi level, we determine the hole surface density to be $p \approx 7.1 \times 10^{12}$ cm$^{-2}$ (*29*)

High-resolution spectroscopic measurements provide the first indication that both single-particle effects and electron-electron interactions break the six-fold symmetry of the hole LLs. Evidence of symmetry breaking can be seen in Fig. 1E, which shows the field evolution of the conductance spectra in one region of the sample where the $N_h = 3,4,5$ LLs are each split into two peaks with different amplitudes—indicating a lifting of the six-fold valley degeneracy of each level to form two- and four-fold degenerate LLs. The fact that the splitting (characterized by a gap $\Delta_{strain}$) occurs away from the Fermi level indicates that it is a single-particle effect. The very weak dependence of $\Delta_{strain}$ on magnetic field and orbital index and the fact that we observe different magnitude gaps in different regions of the sample suggest that local strain underlies this partial symmetry breaking (*29*). As an illustration of the spatial dependence of this behavior, we show in Fig. 1F a spectroscopic line cut from a region of the sample in which the six-fold degeneracy of the $N_h = 3$ LL is lifted to produce either two or three broken-symmetry states, depending on location within the sample.

Electron-electron interactions further lift the LL degeneracy and are manifested in spectroscopic measurements by the appearance of energy gaps when the LLs cross the Fermi level. Figure 1G shows a high resolution measurement of the Fermi level crossing of the $N_h = 4$ LL (in the same area as in Fig. 1E), where over a range of 0.5 T, the four-fold degenerate peak develops an exchange energy gap ($\Delta_{exch} = 450$ µeV) that is coincident with the Fermi level. Although there are spatial variations in the exact magnitude of the gaps between the broken-symmetry LLs, exchange interactions consistently enhance gaps between LLs that are already split by strain and induce a gap between previously degenerate levels when they cross the Fermi level. The magnitude of the exchange gap is consistent with that estimated theoretically for the hole pockets of Bi(111), and it is not related to an Efros-Shlovskii Coulomb gap (*29*)These observations demonstrate that a combination of a single particle effect, likely strain, and many-body interactions lift the six-fold valley degeneracy of the hole LL to produce three broken-symmetry states.

We perform spectroscopic mapping with the STM to directly visualize the underlying quantum Hall wavefunctions and to demonstrate the breaking of crystalline symmetry in these phases. Conductance maps at energies corresponding to each of the three broken-symmetry hole LLs show anisotropic ellipse-like features that point along high-symmetry crystal axes, with relative angles rotated by 120° with respect to each other (Figs. 2A-D). The elliptical features are centered on atomic scale surface defects, and the same defects produce rings in all three directions. This suggests that ellipse orientation is not associated with symmetry breaking from the defect itself, which is further confirmed by atomic resolution topographs (*29*). As we show below, the three different directionalities arise from cyclotron orbits in pairs of hole valleys that are elongated in the same direction. More importantly, such spatially resolved measurements enable us to directly visualize the spontaneous breaking of the LL degeneracy by electron-electron interactions. By tuning the magnetic field to adjust the occupancy of two of the three broken symmetry states, we can contrast spatial maps of the LLs with and without exchange splitting. The measurements in Figs. 2E-G, obtained in the same region as those in Figs. 2A-D, show that the elliptical features in the conductance maps can occur as a superposition of two different orientations, indicating that the symmetry between these two orientations is not broken in the absence of an exchange gap. Contrasting Fig. 2F with Figs. 2B and 2C clearly shows that unidirectional elliptical features emerge as the exchange gap opens, providing a direct and dramatic manifestation of nematic valley-polarized states on the Bi(111) surface.

Another key feature of a nematic electronic phase without long-range order is the presence of domains, which we observe in our system by performing spatially resolved spectroscopy with the STM. We find that the sequence in energy of the three broken symmetry hole LL states can change depending on the location within the sample. An example of this behavior can be seen by contrasting the spectrum and corresponding conductance maps in Figs. 2A-D to those measured about a micron away, shown in Figs. 2H-K. These data reveal that the orientations of the two broken-symmetry states corresponding to the first two peaks in the spectra have switched between the two locations on the Bi surface. Thus, our STM measurements not only show that electron-electron interactions drive nematic behavior, but also illustrate the formation of local nematic domains.

We show below that the elliptical features in our STM conductance maps arise from cyclotron orbit wavefunctions of the broken-symmetry quantum Hall phases that are pinned by surface defects. To characterize these features in detail, we study them in an area with few surface defects (box in Fig. 1A), and examine their dependence on orbital index at a constant magnetic field (14 T) around the same defects (circled in Fig. 1A). The conductance maps shown in Figs. 3A-E are obtained at the energies of the strain-induced broken-symmetry LLs for $N_h = 0$-4, and they reveal concentric ellipses of suppressed conductance similar to those in Fig. 2, with a consistent orientation for all the orbital indices. As orbital index increases, the size of the outermost ring increases, as does the number of concentric rings of suppressed conductance. Around these same surface defects, we observe approximately circular rings in conductance maps measured at the nearby electron LL peak (Fig. 3F), which further confirms that the defects themselves do not break rotational symmetry.

The rings of suppressed conductance for both electron and hole LLs can be understood as a consequence of cyclotron orbits that are shifted in energy because of the sharp potential produced by the atomic surface defects. In the symmetric gauge, the cyclotron orbits of each LL can be labeled by a second orbital quantum number $m$ (*31, 32*). Only the $m = N$ cyclotron orbit has weight at the defect, so it is the only state whose energy is shifted by the defect potential, which we model as a delta function (*29*). Without the defect, conductance maps measured at the LL peak would include DOS contributions from all cyclotron orbits and no spatial variation would be expected. However, because the $m = N$ orbit is shifted to a different energy by the defect, it becomes visible as a decreased conductance in the shape of the wavefunction when measurements are performed at the unperturbed LL energy.

A theoretical model of cyclotron orbit wavefunctions for the surface states of Bi(111) can be used to capture the elliptical features in the STM conductance maps near indivual defects with excellent accuracy. The anisotropy of the surface state hole pockets is reflected in their cyclotron orbit wavefunction, as exemplified by the $m = N_h = 4$ state, whose amplitude $2\pi l_B^2 |\varphi_{4,4}(\vec{r})|^2$ is plotted in Fig. 3G ($l_B = \sqrt{\hbar/eB}$ is the magnetic length). The number of elliptical features in these wavefunctions increases with orbital index and is a reflection of the spatial oscillations of the $m = N_h$ wavefunction, which is proportional to a Laguerre polynomial with $N_h + 1$ peaks (*29*). Using the defects marked in Fig. 1A as the centers of such cyclotron orbits, we simulate the expected conductance pattern by subtracting $2\pi l_B^2 |\varphi_{N,N}(\vec{r})|^2$ from a uniform background (Figs. 3H-K). The similarity with the experimental data in Figs. 3B-E for different $N_h$ states is remarkable, especially given that the only adjustable fit parameter is the anisotropy of the hole pocket effective mass. We extract a ratio of 5 for the hole pocket anisotropy, in good agreement with previous ARPES measurements (*23, 26, 27*) and calculations (*33*). Our model also captures the field dependence of the cyclotron orbit size of the $N_h = 4$ hole LL, as well as that of the electron LLs near the Fermi level. Figure 3L shows the experimentally measured size of the outermost rings for both sets of orbits. They follow the expected $1/\sqrt{B}$ or $1/B$ scaling for hole and electron LLs, respectively, which reflects the dependence of the cyclotron orbit wavefunctions on magnetic length and orbital index (*29*).

Finally, based on the model described above, we anticipate that the suppression we have detected in the conductance maps at the LL peaks should be accompanied by an enhanced conductance relative to the background at other energies. An example of such contrast reversal is shown in Figs. 4A-I, which display conductance maps near an isolated defect over a range of energies within one broken-symmetry $N_h = 4$ LL peak (Fig. 4J). The maps measured at the LL peak and at higher energies show ellipses of suppressed conductance that correspond to a missing cyclotron orbit, whereas at lower energies, such maps show ellipses of higher conductance that indicate the lower energy to which this orbit has been shifted by the defect potential. This reversal of the contrast is clearly illustrated by the energy-averaged line cuts shown in Fig. 4K, which demonstrate that the cyclotron orbit energy has been lowered by about 300 µV by this particular defect. Examining different defects, we have found evidence for both attractive and repulsive potentials from the contrast reversal in the conductance maps (*29*).

Our measurements are in the clean regime where signatures of isolated cyclotron orbits are visible around individual defects, which should be contrasted with previous studies of DOS modulations from drift states moving along equipotential lines in the disordered limit (*34-36*). Cyclotron orbits that are shifted in energy by an isolated defect have been explored in graphene (*32*), and other measurements have indirectly probed the size and shape of cyclotron orbits (*36-38*) by examining LL spatial dependence caused by potential modulations. Here we perform direct two-dimensional mapping of isolated cyclotron orbits, which allows us to visualize nematic order on the Bi(111) surface, where the anisotropic hole mass leads to anisotropic cyclotron orbits.

The Bi(111) 2DEG represents an interesting venue to explore electron-electron interactions within anisotropic valleys. The ability to bring the lowest hole-like LL to the Fermi level, either by external gating or doping, may allow for direct visualization of fractional quantum Hall states and Wigner crystallization with a STM. In addition, the boundaries between different nematic domains are expected to harbor low-energy edge modes that are analogous to topologically protected states (*13*). The ability to generate a valley-polarized nematic phase that can be externally tuned with strain make Bi(111) surface states ideally suited for controlled engineering of anisotropic physical properties. The predicted semimetal-to-semiconductor transition with decreasing thickness in bulk Bi (*18*) means that the transport properties of thin Bi(111) crystals will be dominated by the surface states, yielding further prospects for integration into devices that exploit the unique physical properties reported here.

**Acknowledgments:** We acknowledge helpful discussions with D. A. Abanin, S. A. Kivelson, S. A. Parameswaran, S. L. Sondhi, and A. Yacoby. Work at Princeton has been supported by Gordon and Betty Moore Foundation as part of EPiQS initiative (GBMF4530) and DOE-BES. . This project was also made possible using the facilities at Princeton Nanoscale Microscopy Laboratory supported by grants through, NSF-DMR-1104612, through NSF-MRSEC programs through the Princeton Center for Complex Materials DMR-1420541, LPS and ARO-W911NF-1-0606, ARO-MURI program W911NF-12-1-0461, and Eric and Wendy Schmidt Transformative Technology Fund at Princeton. BEF acknowledges support from the Dicke fellowship. MTR acknowledges support from the NSF Graduate Research Fellowship Program. FW and AHM were supported by DOE Division of Materials Sciences and Engineering grant DE-FG03-02ER45958 and by Welch foundation grant F1473.

**Figure captions**

**Fig. 1**. Landau levels (LLs) of the Bi(111) surface states. (**A**) A typical cleaved Bi(111) surface, with crystallographic axes labeled. The data in Figs. 1E and 1G are an average of spectra measured along the blue line, and the conductance maps in Fig. 3 were performed in the area denoted by the black box. Surface defects are circled in purple, and the inset shows a zoom-in on one defect (inset *z* height scale: 1.3 Å). (**B**) Conductance $G$ as a function of energy $E$ at magnetic field $B = 0$ (blue) and at 14 T (red). The curves are offset (by 0.5) for clarity. At $B = 0$, the data

are taken at temperature $T \approx 4$ K. All other data throughout the manuscript are measured at 250 mK. Inset: Diagram of the Bi(111) first Brillouin zone showing the electron (purple) and hole (blue) Fermi pockets of the surface states. (**C**) Landau fan diagram of $G(E, B)$ that shows crossing electron- and hole-like LLs. The data are averaged over a 20 nm line, with individual spectra showing almost no spatial variation on this energy scale. Select orbital indices $N_e$ and $N_h$ of the respective electron and hole LLs are labeled. (**D**) Higher energy resolution measurement of $G(E, B)$ that clearly shows Fermi level pinning of each hole LL. (**E**) High-resolution measurement of $G(E, B)$ in a region where the $N_h = 3,4,5$ LLs each show splitting into a two-fold degenerate and a four-fold degenerate LL peak. Data are averaged over the blue line in Fig. 1A. (**F**) Linecut of spectra showing strain-induced splitting of the sixfold degenerate $N_h = 3$ LL into two or three peaks, depending on position. Numbers in parentheses denote the degeneracy of each broken-symmetry state. (**G**) Zoom-in on $G(E, B)$ in the same location as in A. The four-fold degenerate peak further splits into two distinct LLs as it crosses the Fermi level, indicating broken symmetry states arising from exchange interactions. Arrows mark $\Delta_{strain}$ and $\Delta_{exch}$.

**Fig. 2**. Rotational symmetry breaking and local domains of a nematic electronic phase. (**A**) Average conductance spectrum at 12.9 T, measured over a 100 nm linecut (which exhibits little spatial dependence) near the start of the linecut in Fig. 1F, showing three broken-symmetry hole LLs, two of which are split by exchange interactions at the Fermi level. (**B**)-(**D**) Spatial maps of $G/\bar{G}$ at energies corresponding to the three split hole LL peaks. Ellipses of reduced conductance are centered on surface defects, with different orientations at each energy. (**E**) Average conductance spectrum at 14 T, measured in the same location as in A. The spectrum shows restored symmetry of the exchange-split LLs in A to produce a four-fold degenerate LL. (**F**)

Spatial map of $G/\bar{G}$ at the energy of the four-fold degenerate LL peak which shows ellipses with two orientations. (**G**) Spatial map of $G/\bar{G}$ at the energy of the two-fold degenerate LL peak that is split from the four-fold degenerate peak by strain, showing the same unidirectional behavior as in D. The spatial maps in panels F and G are measured in the same area as panels B-D. (**H**) Average conductance spectrum (measured over a 100 nm linecut that exhibits little spatial dependence) at 12.9 T in a location about 1 micron away from the region shown in (A-D). (**I**)-(**K**) Spatial maps of $G/\bar{G}$ in the new location at energies corresponding to the three split hole LL peaks. The energetic order of the three directions is different, with the first two orientations switched, demonstrating the presence of domains. For all conductance spectra, the electron LLs are labeled, and the hole LL degeneracy is denoted in parentheses near each peak.

**Fig. 3**. Isolated anisotropic cyclotron orbits and theoretical modeling. (**A**)-(**E**) Spatial maps of $G/\bar{G}$ at 14 T in the area denoted by the black box in Fig. 1A, at energies corresponding to the strain-induced broken-symmetry hole LL for orbital indices $N_h = 0$-4. Isolated anisotropic cyclotron orbits are present around surface defects. (**F**) Spatial map of $G/\bar{G}$ in the same area at the energy of the $N_e = 8$ LL, showing circular rings of suppressed conductance that occur around the same surface defects (black arrows). The weak elliptical feature around the lower defect is related to a missing cyclotron orbit from the $N_h = 3$ LL at a nearby energy. The trapezoidal feature in the background conductance results from the shape of the terrace because the LL visibility is suppressed near step edges. (**G**) Amplitude $2\pi l_B^2 |\varphi_{4,4}(\vec{r})|^2$ of the $m = N_h = 4$ cyclotron orbit wavefunction. (**H**)-(**K**) Simulated maps of the expected conductance, $1 - 2\pi l_B^2 |\varphi_{N,N}(\vec{r})|^2$, with individual cyclotron orbits centered on the surface defects circled in Fig. 1A. The size and shape of the simulated conductance are a good match to the data in panels B-E.

(**L**) Semimajor axis size of the cyclotron orbits for $N_h = 4$ (blue) and ring size of those from electron LLs near the Fermi level (red) as a function of magnetic field. Dashed lines are fits to the field dependence of the extracted sizes.

**Fig. 4**. Energy shift of the cyclotron orbits. (**A**)-(**I**) Spatial maps of $G/\bar{G}$ around an isolated impurity at $B = 10$ T with energy spaced by 100 µeV throughout one broken-symmetry $N_h = 4$ LL peak. These maps show the shift to lower energy of the $m = N$ cyclotron orbit. (**J**) Corresponding conductance spectrum (averaged over a 12 nm x 2.5 nm area centered about 5 nm underneath the defect) marked with colored circles for each mapped energy. (**K**) Oscillations of $G/\bar{G}$ along the semiminor axis, averaged over 100 and 200 µeV (blue), and 400 and 500 µeV (red), respectively, highlighting the constrast reversal in the maps.

**Supplementary Materials**

Materials and Methods

Supplementary Text

Figs. S1 to S8

Tables S1 and S2

References (39-43)

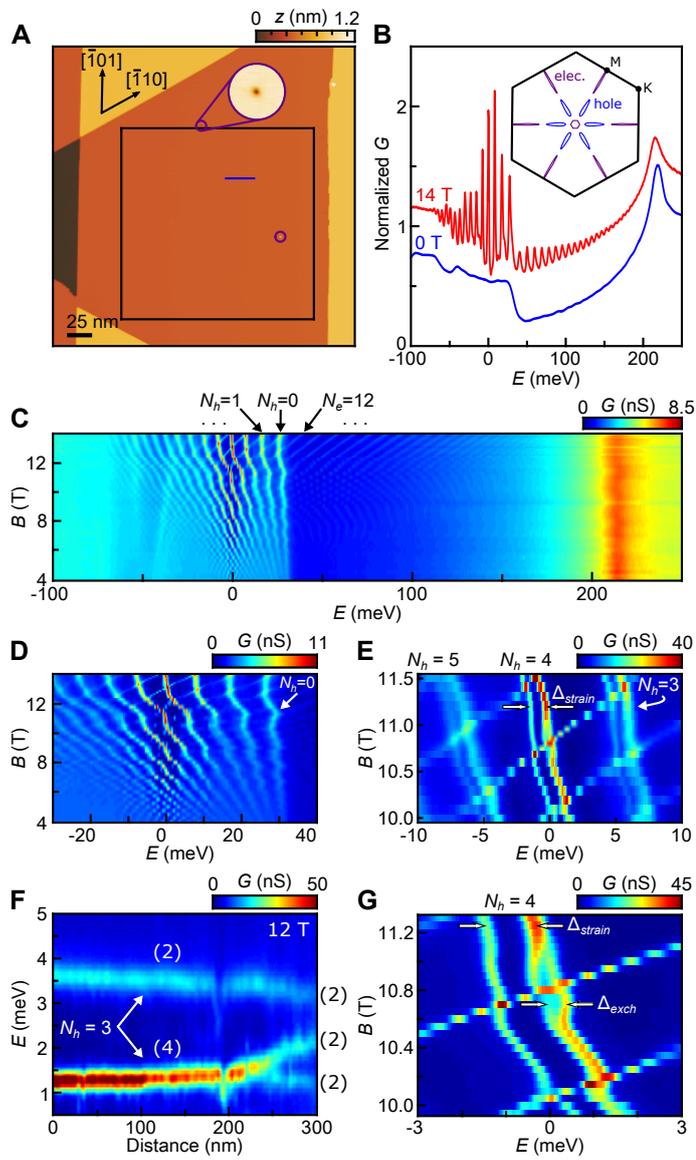

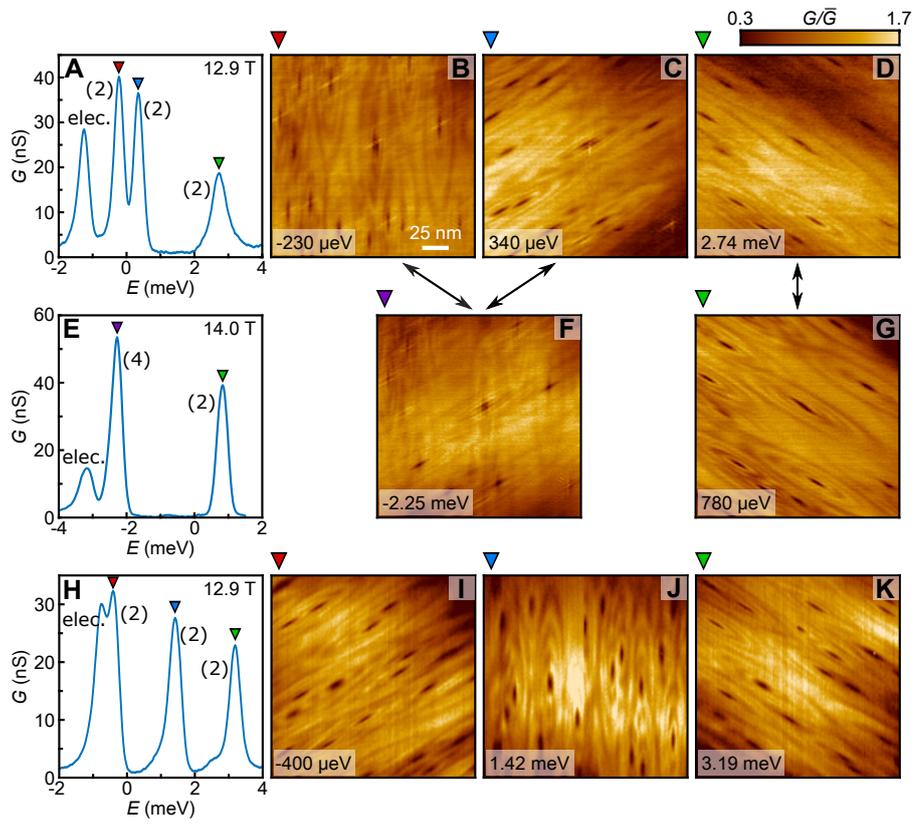

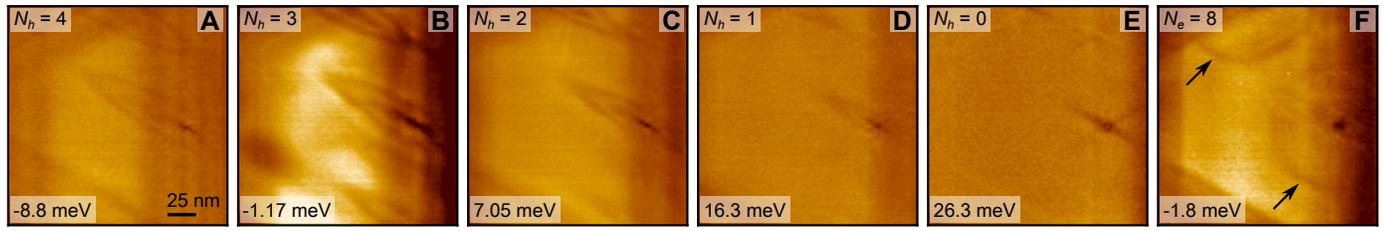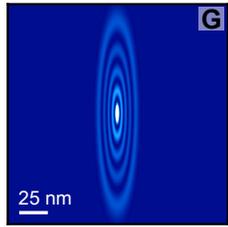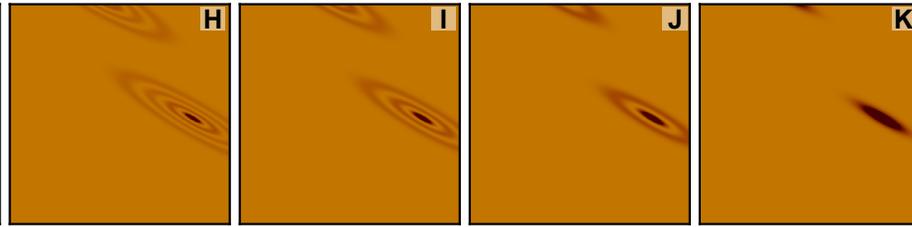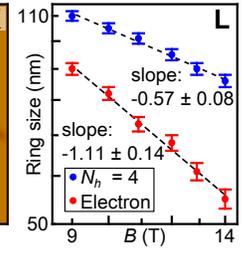

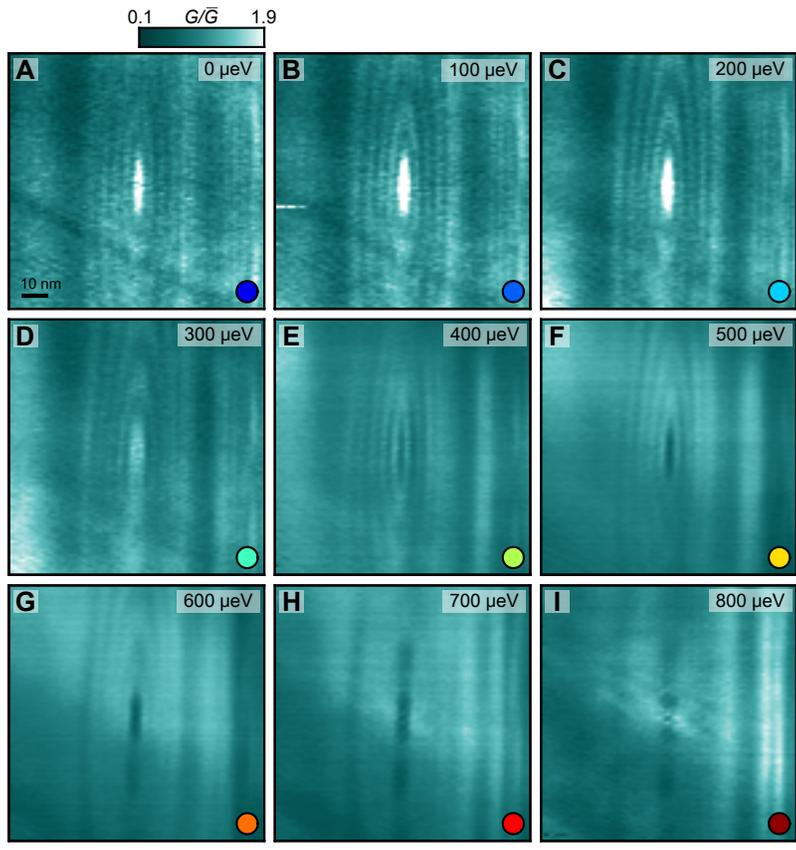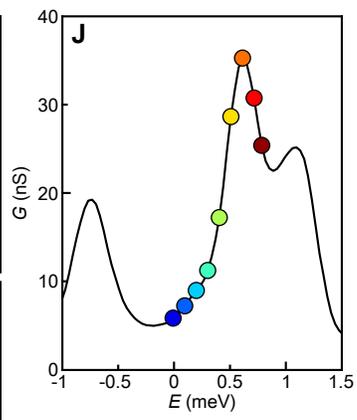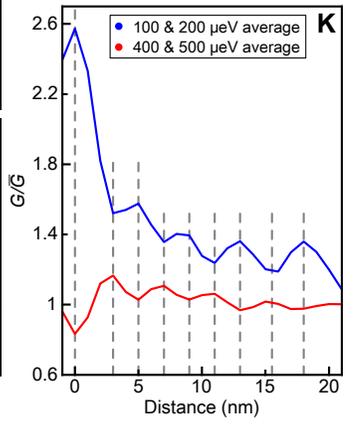

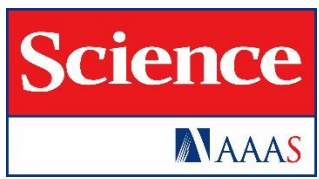

# Supplementary Materials for

Observation of a Nematic Quantum Hall Liquid on the Surface of Bismuth

Benjamin E. Feldman[1†], Mallika T. Randeria[1†], András Gyenis[1†], Fengcheng Wu[2], Huiwen Ji[3], R. J. Cava[3], Allan H. MacDonald[2], and Ali Yazdani[1]*

[†]These authors contributed equally to this work.
*correspondence to: yazdani@princeton.edu

**This PDF file includes:**

Materials and Methods
Supplementary Text
Figs. S1 to S8
Tables S1 and S2
References

**Materials and Methods**

Single Bi crystals were grown using the Bridgman method from 99.999% pure Bi that had been treated to remove oxygen impurities. The samples were cleaved in ultrahigh vacuum at room temperature and immediately inserted into a home-built dilution refrigerator STM (*39*) and cooled to cryogenic temperatures. The cleaved Bi(111) surface shows almost no defects at this point, as illustrated by Figs. 1A and 3, and the number of defects is stable at low temperature. Following a thermal cycle to 30 K and back down to 250 mK, a larger number of defects were present, as illustrated by Fig. 2. These are likely hydrogen molecules adsorbed on the surface due to outgassing of the microscope walls. Similar nematic behavior was observed both before and after the thermal cycle.

Except where noted, measurements were performed at 250 mK using a W tip. Spectra and conductance maps were acquired using a lock-in amplifier at a frequency of 700.7 Hz and with AC rms excitation $V_{rms}$ varying from 30 µV to 3 mV. Except where noted, the setpoint voltage was $V_{set} = -400$ mV and the setpoint current was $I_{set} = 5$ nA. We detail the measurement parameters for each figure below.

Fig. 1: $I_{set} = 30$ pA for panel A, 100 pA for the zero field data in panel B, and 2 nA for all other measurements. $V_{rms} = 3$ mV for the zero field data in panel B, 500 µV for panel C and the 14 T data in panel B, 250 µV for panel D, 100 µV for panel E, 74 µV for panel F, and 30 µV for panel G.

Fig. 2: $V_{rms} = 30$ µV for panels A, E, and H; $V_{rms} = 74$ µV for panels B-D, F, G, and I-K.

Fig. 3: $V_{rms} = 74$ µV in panels A-F.

Fig. 4: $V_{rms} = 74$ µV in all panels.

**Supplementary Text**

Bi(111) surface state band structure

As a reference and for completeness, we plot in Fig. S1 the Bi(111) surface state band structure along the high-symmetry directions (adapted from Ref. 24). There are two spin-orbit split bands, denoted by the purple and blue lines, which give rise to the surface state electron and hole pockets, respectively (Fig. 1B, inset). The top of the higher energy (purple) band along the Γ-M direction gives rise to the sharp peak in zero-field conductance that we observe at 220 meV, while the drop in conductance near 33 meV corresponds to the upper edge of the lower energy (blue) band (Fig. 1B).

Electron and hole surface densities, and charge redistribution between pockets

The hole density of the Bi(111) surface states can be determined by the magnetic field at which each respective hole LL crosses the Fermi level (Table S1). At high magnetic field, where the LLs are pinned to the Fermi level for an extended field range, we use the midpoint as the crossing point, and take the LL to be half-filled at this point. Strictly speaking, this introduces a small error because filling factor $v$ and magnetic field are inversely related, so the midpoint in field is not the midpoint in filling factor. Nonetheless, this error is small, especially at low fields where pinning is not observed over an appreciable field range. Each LL crossing provides an independent measure of the total hole density $p = veB/h$, and all derived values agree to within 2%, yielding an average of $(7.08 \pm 0.02) \times 10^{12}$ cm$^{-2}$. This consistency confirms that the surface hole density is fixed in our system, except for the variations described at the end of this section.

In principle, a similar procedure can be used to determine the surface electron density $n$. However, the electron case is more complicated because LLs could arise from the central hexagonal pocket and/or the three elongated pockets around the M points. Theoretical calculations (*26, 33*) show that the pockets around the M points are actually surface resonances due to strong hybridization with the bulk, suggesting our surface measurement is most sensitive to the pocket around the Γ point. This is confirmed by the circular electron cyclotron orbits that we observe in Fig. 3F; the oblong pockets around the M points would give rise to anisotropic rings similar to the hole states. We do not observe any LLs from the electron pockets around the M points, so we can only determine the surface electron density arising from the central pocket.

The lowest electron LL is not visible in our data, which is likely due to a combination lifetime broadening away from the Fermi level and surface-bulk hybridization caused by overlap of bulk states with the bottom of the surface band that gives rise to the central electron pocket (*24*). Therefore, we cannot directly determine the orbital index $N_e$ of each LL, and the filling factor needs to be assigned using a different method. We try three different filling factor assignments and calculate the corresponding density for each case according to $n = veB/h$ (Table S2). Only for the case where the electron LL crossing at 13.1 T corresponds to $v = 9.5$ do we find a constant carrier density from the central electron pocket of $n = (3.02 \pm 0.02) \times 10^{12}$ cm$^{-2}$. For the cases where filling factor is higher or lower, the derived density either steadily increases or steadily decreases with field, inconsistent with our assumption of constant surface electron density. We therefore take the first case to be the correct assignment. Both this electron density and the derived hole density above are consistent with previous measurements of Bi(111) thin films (*28*).

Although the electron and hole carrier densities that we measure are consistent across a wide field range, we observe temporary charge redistribution between pockets as the electron and hole LLs cross at the Fermi level. The evidence for this comes from the varying behavior of the electron LLs near the Fermi level (Fig. S2). When there are no other states nearby (*e.g.* around 11.8 T), the electron LL is pinned to the Fermi level over an appreciable field range, indicating that the Fermi level is set by the chemical potential of the electrons. In contrast, when a hole LL is pinned to the Fermi level, the electron LLs cross through it without any sign of pinning. At 10.9 T, the $N_e = 11$ LL is about 1 mV above the Fermi level and is therefore empty. By 10.7 T, it has already crossed through the $N_h = 4$ LL to negative energy (-1 mV) and is therefore completely filled. This corresponds to an increase in electron density from $2.90x10^{12}$ cm$^{-2}$ to $3.10x10^{12}$ cm$^{-2}$. In this regime, the Fermi level is primarily controlled by the chemical potential of the hole LL.

The difference in behavior signifies an interesting competition between carriers from the electron and hole pockets, and it can be understood by comparing the energetics of electron and hole LLs as a function of magnetic field. As magnetic field is decreased, the energies of the electron LLs decrease. In contrast, the energies of hole LLs increase as magnetic field is lowered (this corresponds to a lower energy for holes, but a higher energy for electrons). Therefore, it is energetically favorable for electrons to preferentially occupy states in the hole pockets at magnetic fields above crossings between electron and hole LLs, whereas it becomes more favorable to occupy states in the electron pockets below the crossing. As magnetic field is decreased through a crossing, charge is transferred from the hole pocket to the electron pocket, effectively increasing both the electron and hole filling factors by one. The amount of charge transfer expected corresponds to one electron LL, or about $2.6x10^{11}$ cm$^{-2}$ at 10.9 T, similar to our observed change in electron density. We remark that the charge transferred reaches 5-10% of the total charge present in the central electron pocket. This is already substantial, and the charge transfer will be even larger at higher magnetic field. To the best of our knowledge, spontaneous charge transfer between electron and hole pockets has never been observed.

We note that the behavior described above involves charge transfer between pockets in k-space, and occurs in a spatially localized manner; it does not imply charge transfer within the sample in real space. Our measurements probe this phenomenon locally, and a measurement at a single location does not imply global effects or uniformity, but we have observed similar behavior in multiple positions.

Field and orbital dependence of $\Delta_{strain}$ and evidence that it is caused by local strain

Figure S3A shows the magnitude of $\Delta_{strain}$ for $N_h = 3-5$ derived from Fig. 1E. All the gaps cluster together, showing that there is no appreciable dependence on orbital index. We do not observe a strong dependence of $\Delta_{strain}$ on magnetic field either, beyond the exchange enhancement at the Fermi level for $N_h = 4$ (more clearly visible in Fig. S3B, which is obtained from the data in Fig. 1G). Variations as a function of position are larger than variations as a function of magnetic field, supporting our conclusion that the main source of the single-particle splitting is likely local strain. Moreover, it is reasonable to expect that some amount of strain is present in our sample as a result of the difference in thermal contraction of the sample relative to its substrate and/or the presence of twin boundaries, which have been observed in bulk Bi (*40*). Strain has been shown to couple to the valley degree of freedom and even lead to full valley polarization in AlAs (*16, 41*). Thus, it is logical to expect strain to play a role in the single-particle valley splitting that we observe. See also the discussion regarding the domain wall and nearby strain defect below for further evidence that strain affects the energies of the hole LLs.

Efros-Shlovskii Coulomb Gap

A soft gap in the density of states (DOS) near the Fermi level, known as the Efros-Shlovskii (Coulomb) gap, can result from the combination of electron localization and Coulomb interactions (*30, 42*). The energy gap that we observe as hole LLs cross the Fermi level is clearly not an Efros-Shlovskii gap, because we do not observe similar behavior when the singly degenerate electron LLs cross the Fermi level (see, *e.g.*, Fig. S2). A weak suppression of the DOS is visible for energy $|E| <$ 3-5 meV in Figs. 1D and 1E. This effect, whose energy scale is about ten times that of the exchange splitting, could result in part from a Coulomb gap, and could also reflect a low DOS between LLs that is especially pronounced near the Fermi level due to reduced lifetime broadening.

Nematic behavior in a second location

The data presented in Fig. 1G of the main text show the field dependence of the energy gaps between broken-symmetry states at one location on the sample surface. As described in the main text, we observe three broken-symmetry states, two of which are split by exchange interactions as the otherwise four-fold degenerate hole LL crosses the Fermi level. A spectrum measured in this location at $B$ = 10.9 T is shown in Fig. S4A, and the corresponding conductance maps at each broken-symmetry LL peak are shown in Figs. S4B-D. Similar to the data presented in Figs. 2A-D of the main text, each broken-symmetry LL peak exhibits ellipses with different directionalities. We observe only minimal splitting of these states when they are not near the

Fermi level. Thus, this location provides a second example where electron-electron interactions at the Fermi level play a dominant role in the formation of nematic electronic order. We also note that the energetic order of directionalities of the three broken-symmetry states is different from that in Figs. 2A-D, another example of a local nematic domain. For completeness, we also include a conductance map measured at the nearby electron LL peak (Fig. S4E), which shows approximately circular rings, similar to Fig. 3F in the main text.

Theoretical calculation of cyclotron orbits of Bi(111) hole states

To calculate the cyclotron orbits that arise from the anisotropic hole pockets of the Bi(111) surface, we assume a parabolic dispersion and approximate the pockets as ellipses. This yields a Hamiltonian for a single pocket

$$H = -\frac{\pi_\parallel^2}{2m_\parallel} - \frac{\pi_\perp^2}{2m_\perp} + E_0, \qquad (S1)$$

where $E_0 \approx 33$ meV is the energy of the hole band edge, $\pi$ is the momentum, and $m$ is the effective mass, with $\parallel$ and $\perp$ denoting the semimajor and semiminor ellipse axes, respectively. We rescale the coordinates

$$X = \sqrt[4]{m_\parallel/m_\perp}\, x, \; \Pi_x = \sqrt[4]{m_\perp/m_\parallel}\, \pi_x, \qquad (S2)$$

$$Y = \sqrt[4]{m_\perp/m_\parallel}\, y, \; \Pi_y = \sqrt[4]{m_\parallel/m_\perp}\, \pi_y, \qquad (S3)$$

to generate an isotropic Hamiltonian

$$H = -\frac{\Pi^2}{2M} + E_0, \qquad (S4)$$

where $M = \sqrt{m_\parallel m_\perp}$. This allows us to write down the corresponding cyclotron orbit wavefunctions in the symmetric gauge by analogy to conventional LLs in isotropic systems,

$$\varphi_{N,m}(Z) = \frac{1}{\sqrt{2\pi l_B^2}} \left(\frac{N!}{m!}\right)^{\frac{1}{2}} \left(\frac{Z^*}{\sqrt{2}}\right)^{m-N} e^{\frac{-|Z|^2}{4}} L_N^{m-N}\left(\frac{|Z|^2}{2}\right), m \geq N$$

$$\varphi_{N,m}(Z) = \frac{1}{\sqrt{2\pi l_B^2}} \left(\frac{m!}{N!}\right)^{\frac{1}{2}} \left(\frac{Z}{\sqrt{2}}\right)^{N-m} e^{\frac{-|Z|^2}{4}} L_m^{N-m}\left(\frac{|Z|^2}{2}\right), m < N$$

(S5)

Here, $Z = (X + iY)/l_B$ where $l_B = \sqrt{\hbar/eB}$ is the magnetic length, $N$ is the orbital index, $m = 0, 1, 2, \ldots$ is a quantum number relating to the angular momentum of the cyclotron orbit relative to the origin, and $L_N$ is a Laguerre polynomial. We note that $\varphi_{N,m}$ has a non-zero value at the origin only if $m = N$, in which case the cyclotron orbit wavefunction can be written as

$$\varphi_{N,N}(Z) = \frac{1}{\sqrt{2\pi l_B^2}} e^{\frac{-|Z|^2}{4}} L_N\left(\frac{|Z|^2}{2}\right). \tag{S6}$$

The above simplification is important when we consider the effect of the sample defects on individual cyclotron orbits, as detailed below. Due to the atomic length scale of the surface defects that we observe and the relatively strong screening in bismuth, we model the defect potential by a δ-function:

$$U = \alpha \delta(\vec{r}), \tag{S7}$$

where α is the strength of the defect potential. The energy shift of the $\varphi_{N,N}$ state relative to the $\varphi_{N,m \neq N}$ states is

$$E_{N,N} - E_{N,m \neq N} = \alpha |\varphi_{N,N}(0)|^2 = \frac{\alpha}{2\pi l_B^2}, \tag{S8}$$

assuming that this shift is small compared to the LL spacing. Therefore, the local density of states is proportional to $|\varphi_{N,N}(\vec{r})|^2$ for measurements performed at energy $E_{N,N}$, whereas it is proportional to

$$\sum_{m \neq N} |\varphi_{N,m}(\vec{r})|^2 = \frac{1}{2\pi l_B^2} - |\varphi_{N,N}(\vec{r})|^2 \tag{S9}$$

for measurements at $E_{N,m \neq N}$. In the latter case, which corresponds to the experimental measurements in Figs. 2 and 3, peaks in $|\varphi_{N,N}(\vec{r})|^2$ therefore correspond to local minima in conductance, as we observe.

Theoretical estimates of exchange splitting

We outline below a calculation of the exchange splitting of the hole LLs on the Bi(111) surface. We note that the spin degree of freedom is already fixed by the strong spin-orbit coupling and subsequent splitting of the surface bands so that the only remaining internal degree of freedom subject to exchange is the valley index. The exchange energy is a function of the Coulomb potential as well as the surface state wavefunctions, and we use the wavefunctions derived in the previous section in our calculation below. We start with the equation,

$$\Delta_{exch}(N,B) = \int \frac{d^2\vec{q}}{(2\pi)^2} V(\vec{q}) \mathrm{Exp}\left[-\frac{Q^2 l_B^2}{2}\right] \left| L_N\left(\frac{Q^2 l_B^2}{2}\right) \right|^2, \qquad (S10)$$

where wavevector $\vec{q}$ is related to $\vec{Q}$ by $q_x = \sqrt{\lambda} Q_x$ and $q_y = Q_y/\sqrt{\lambda}$, with $\lambda = \sqrt{m_\parallel/m_\perp} \approx 5$ the aspect ratio of the hole Fermi pocket. We approximate the interaction potential $V(\vec{q})$ by the static screened interaction at zero magnetic field,

$$V(\vec{q}) \approx \frac{2\pi^2}{\varepsilon} \frac{1}{q + 2\pi e^2 v_0/\varepsilon}, \qquad (S11)$$

where $\varepsilon$ is the dielectric constant and $v_0$ is the surface density of states at the Fermi level, which can be extracted from the LL spacing using the semiclassical quantization rule. At $B = 10.9$ T, $2\pi l_B^2 v_0 \approx \frac{1}{7.6 \text{ meV}} + \frac{6}{6.1 \text{ meV}}$, where 7.6 meV and 6.1 meV are the respective LL spacing for electrons and holes, and the factor of 6 in the numerator of the second term takes into account the degeneracy of the hole pockets.

We measure a gap $\Delta_{exch}$ = 450 µeV for the $N_h$ = 4 hole LL at $B$ = 10.9 T (Fig. 1G). To reproduce this number using the above equations, we extract a dielectric constant $\varepsilon$ = 45, which is about half the value in bulk Bi, as expected for a surface state bounded by vacuum on the other side. Using this same dielectric constant, we calculate theoretically expected exchange gaps for the $N_h$ = 3 state at $B$ = 12.9 T and 14 T, which are 560 µeV and 600 µeV, respectively. These numbers closely match the experimentally measured values of 570 µeV and 630 µeV, respectively, which serves as a consistency check for the model.

Cyclotron orbit field dependence

In Fig. 3L, we plot the size of the hole and electron cyclotron orbits as a function of magnetic field. The size of a cyclotron orbit is proportional to $\sqrt{(2N+1)}l_B$, so increasing the magnetic field should lead to smaller rings. For fixed $N$, the cyclotron orbit should scale as $1/\sqrt{B}$, and the fit to our experimentally measured ellipse size for $N_h$ = 4 matches this prediction within the experimental uncertainty. For states at a given constant energy in a system with fixed charge, $N$ should scale as $1/B$, leading to a factor of $1/\sqrt{B}$ change in cyclotron orbit size. This, combined with the changing magnetic length, explains the approximate $1/B$ scaling that we observe for the electron LLs near the Fermi level.

A second example of cyclotron orbit energy dependence

A spectrum at at $B$ = 14 T measured in the same location as in Fig. 3 shows three broken-symmetry $N_h$ = 3 LL peaks (Fig. S5A). A series of conductance maps at energies that span the three broken-symmetry LL peaks is shown in Figs. S5B-P. First, we note that the sequence of orientations of the three broken-symmetry states is identical to that found in the same area for the

$N_h = 4$ state (Fig. S4). In addition, careful examination reveals bright vertical rings surrounding two different spots in Figs. S5N-P, which correspond to subsurface defects that have no topographic signal. In contrast, these same defects give rise to vertical dark rings of decreased conductance in Figs. S5J-L. Thus, for these subsurface defects, we extract an energy shift of the $m = N_h = 3$ cyclotron orbit of about 420 µV. Similar behavior is visible from the same spots for the diagonal rings in Figs. S5G-K, as well as in Figs. S5B-F.

The similar energy shift caused by a given defect on cyclotron orbits from each split hole LL peak provides further confirmation that the defects themselves do not break rotational symmetry. The data also illustrate that different defects produce different effective potentials. In fact, the subsurface defects of Fig. S5 are repulsive because they shift the cyclotron orbits to higher energy, opposite from the behavior observed in Fig. 4 of the main text.

Spectroscopic linecut across a domain wall

We have measured conductance spectra along a line that extends from one nematic domain to another, taken along the path shown in Fig. S6A. At either end of the linecut, the LL energies do not vary strongly with position (Fig. S6B), and these two regions respectively correspond to domains that are contiguous with those presented in Fig. 2. Although the spectroscopic signal remains strong except near the step edges, towards the middle of the linecut, the LL energies rapidly disperse with position and a clear LL crossing is visible about 400 nm from the start of the linecut (Fig. S6B). This is exactly the signature expected from a domain wall.

The domain wall occurs near a pronounced strain defect that is visible in the topograph (arrow in Fig. S6A), and strain may act as a catalyst for the domain wall formation. The dramatic

LL dispersion near this strained region is further evidence that the single-particle LL splitting characterized by $\Delta_{strain}$ results from local strain in the sample.

Atomic resolution image of the Bi(111) lattice and a defect

Figure S7A shows an atomic resolution image of the Bi(111) surface which includes an isolated defect. The topograph, which was taken in the vicinity of the conductance maps in Figs. 2I-K, shows a perfectly ordered lattice, demonstrating the high sample quality. This rules out variations in the vertical confinement of the 2DEG as a possible cause of the symmetry breaking that we observe. In addition, the topograph illustrates that the defect does not break the three-fold rotational symmetry of the lattice.

The corresponding Fourier transform (Fig. S7B) shows sharp Bragg peaks from the bismuth lattice. From the positions of the Bragg peaks, we can extract the lattice constant in each direction. We obtain an in-plane interatomic spacing of approximately 4.67 Å, similar to literature values (*25*), and the lattice constant that we extract varies less than 5% between the three principal crystallographic directions. This variation is within the instrumental uncertainty that arises from a combination of piezoscanner calibration uncertainty as well as drift during measurement. Thus, we cannot conclusively say that we observe a strained lattice in atomic resolution images, though strain may still be present below our detection theshold (this is likely, as argued above).

Temperature and doping dependence of the nematic behavior

In the absence of extrinsic symmetry-breaking terms, a nematic electronic phase is expected to have a critical temperature above which it becomes isotropic. We can gain some insight by comparing our data to previous 4.3 K measurements of a Bi(111) thin film (*28*). The thin film showed similar electron and hole LL energies to our sample, but did not exhibit any broken symmetry in the hole LLs. This implies that either the measurements were performed above the critical temperature, or that thermal broadening made it impossible to resolve an exchange gap associated with nematicity. Regardless, in the absence of splitting, one would not expect to see any rotational symmetry breaking, so the measurements of Ref. 28 suggest that any nematic observable would not survive up to 4.3 K. We can estimate an upper bound for the temperature above which nematicity would disappear based on the magnitude of the exchange splitting that we observe. A gap of 500 µeV becomes equvalent to the thermal broadening $3.5k_BT$ at a temperature $T \sim 1.6$ K. This temperature threshold is consistent with the currently available experimental data.

Another example of an external tuning parameter that can affect nematicity is disorder, which has the further advantage that it is not complicated by thermal broadening. To explore the effect of disorder, we have performed measurements at 250 mK of an intentionally doped Bi sample (Fig. S8A). In these more disordered samples, cyclotron orbits of all three directionalities are visible throughout each LL peak. A representative conductance map taken at the Fermi level at $B = 10$ T is shown in Fig. S8B. Many superimposed cyclotron orbits are visible because so many dopants are present, and the presence of features of all three directionalities demonstrates that disorder destroys the nematic order. Further exploration of the rich phase diagram of broken-symmetry states that are expected as a function of temperature and disorder (*14, 43*) represents an appealing direction for future research.

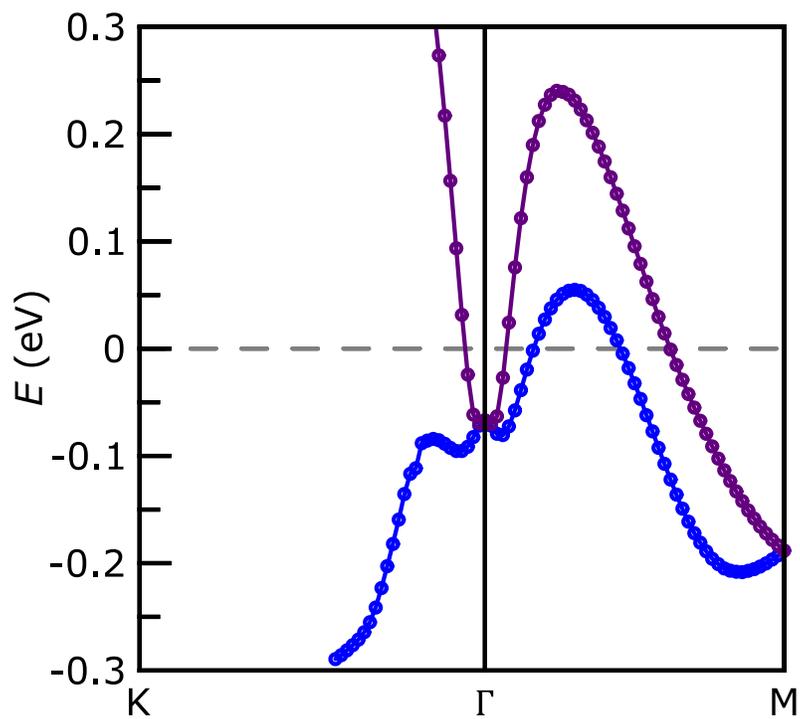

**Fig. S1**. Bi(111) surface state band structure along high-symmetry directions. Strong Rashba spin-orbit coupling gives rise to spin-split bands (blue and purple lines) that exhibit multiple Fermi level crossings to produce multiple electron and hole pockets. Figure adapted from ref. (*22*).

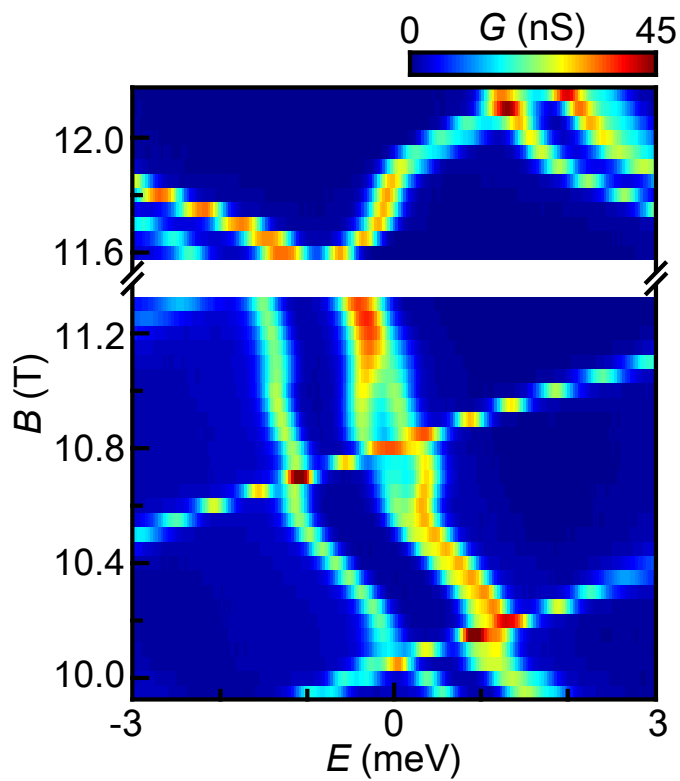

**Fig. S2**. Tunneling conductance $G$ as a function of energy $E$ and magnetic field $B$. Data are averaged over the blue line in Fig. 1A, and the lower field data are the same as those shown in Fig. 1G, but are reproduced here to improve readability. The electron Landau levels (LLs) exhibit two types of behavior as they cross the Fermi level: pinning when there are no proximal hole-like states (e.g. around 11.8 T), or rapid dispersion through hole LLs (e.g. at 10.8 T and 10.1 T). The two different behaviors indicate a competition between electron-and hole-like states in a magnetic field. $V_{rms} = 30$ μV.

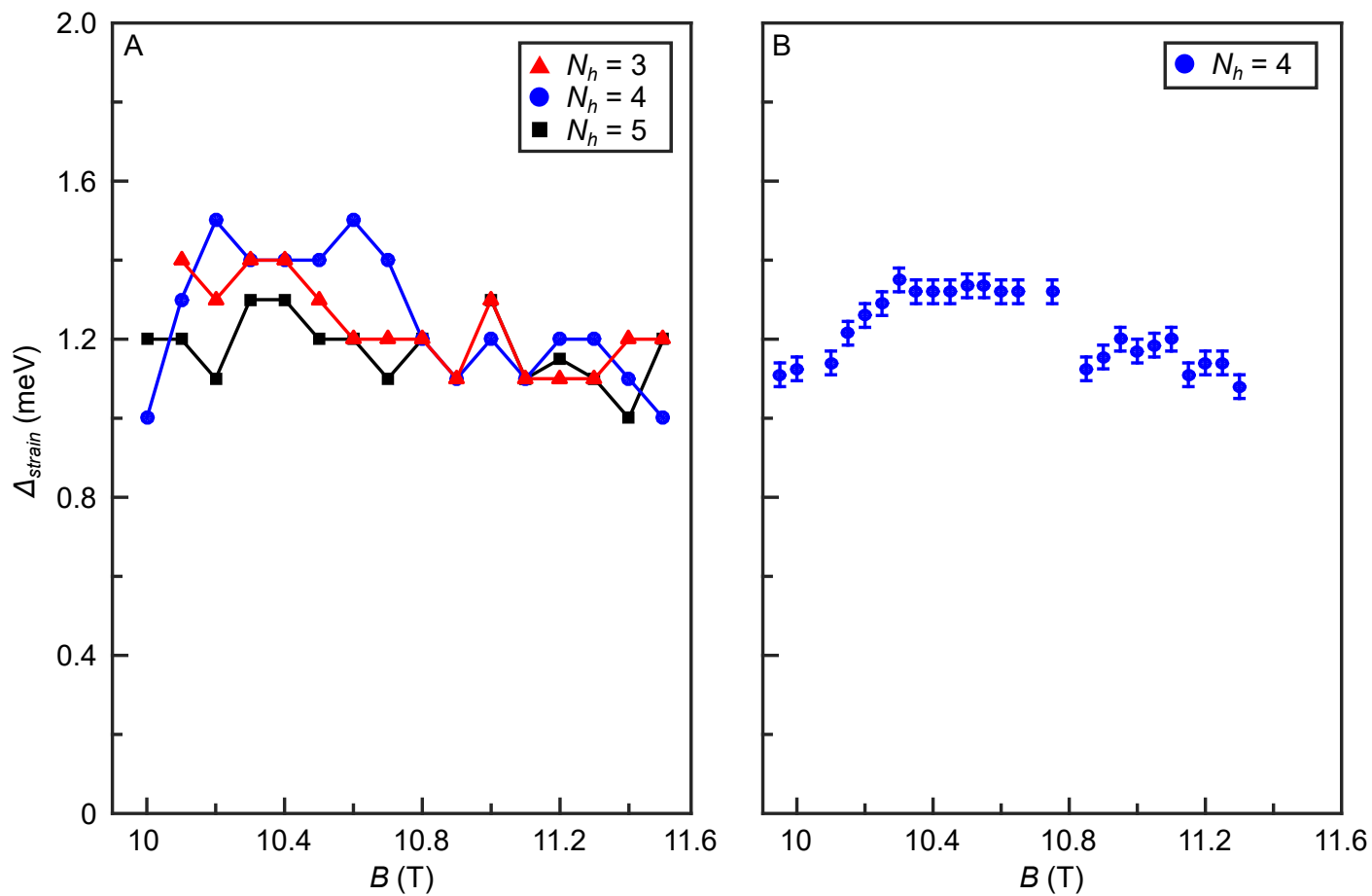

**Fig. S3**. Field and orbital dependence of the gap $\Delta_{strain}$. (**A**) Magnitude of $\Delta_{strain}$ as a function of magnetic field derived from the data in Fig. 1E for the hole LLs with orbital indices $N_h = 3,4,5$. All gaps cluster together and exhibit no appreciable field dependence, except for exchange enhancement of the $N_h = 4$ gap when it coincides with the Fermi level (*i.e.* between 10.1 and 10.7 T). (**B**) Similar data (with error bars) showing $\Delta_{strain}$ for $N_h = 4$ derived from the higher resolution data in Fig. 1G.

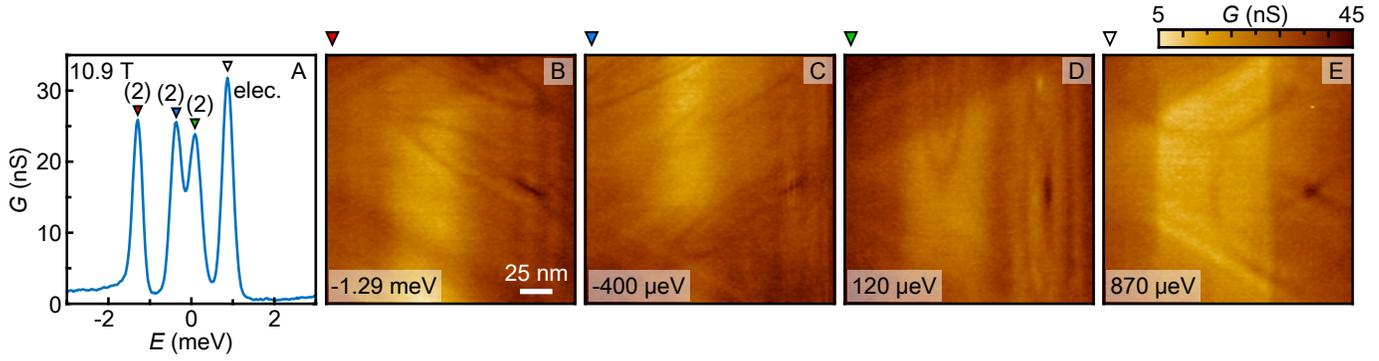

**Fig. S4**. Nematic electronic behavior in a second location. (**A**) Average conductance spectrum measured at $B = 10.9$ T in the same location as the data in Figs. 1E and 1G. The $N_h = 4$ LL is split into three peaks, two of which are split by exchange interactions at the Fermi level. The $N_e = 11$ LL is also visible. The degeneracy of each hole LL is labeled in parentheses. (**B**)-(**D**) Conductance maps measured at the energies of each broken-symmetry hole LL peak. Isolated anisotropic cyclotron orbits are visible around surface defects, with a different orientation for each LL peak. The energetic order of the three directionalities is different from that shown in Figs. 2A-D, indicating a different nematic domain.(**E**) Conductance map measured at the energy of the $N_e = 11$ LL. Approximately circular rings are visible around the same surface defects. Due to the lower magnetic field and larger orbital index, the rings are larger than those in Fig. 3F. $V_{rms} = 30$ μV for all panels.

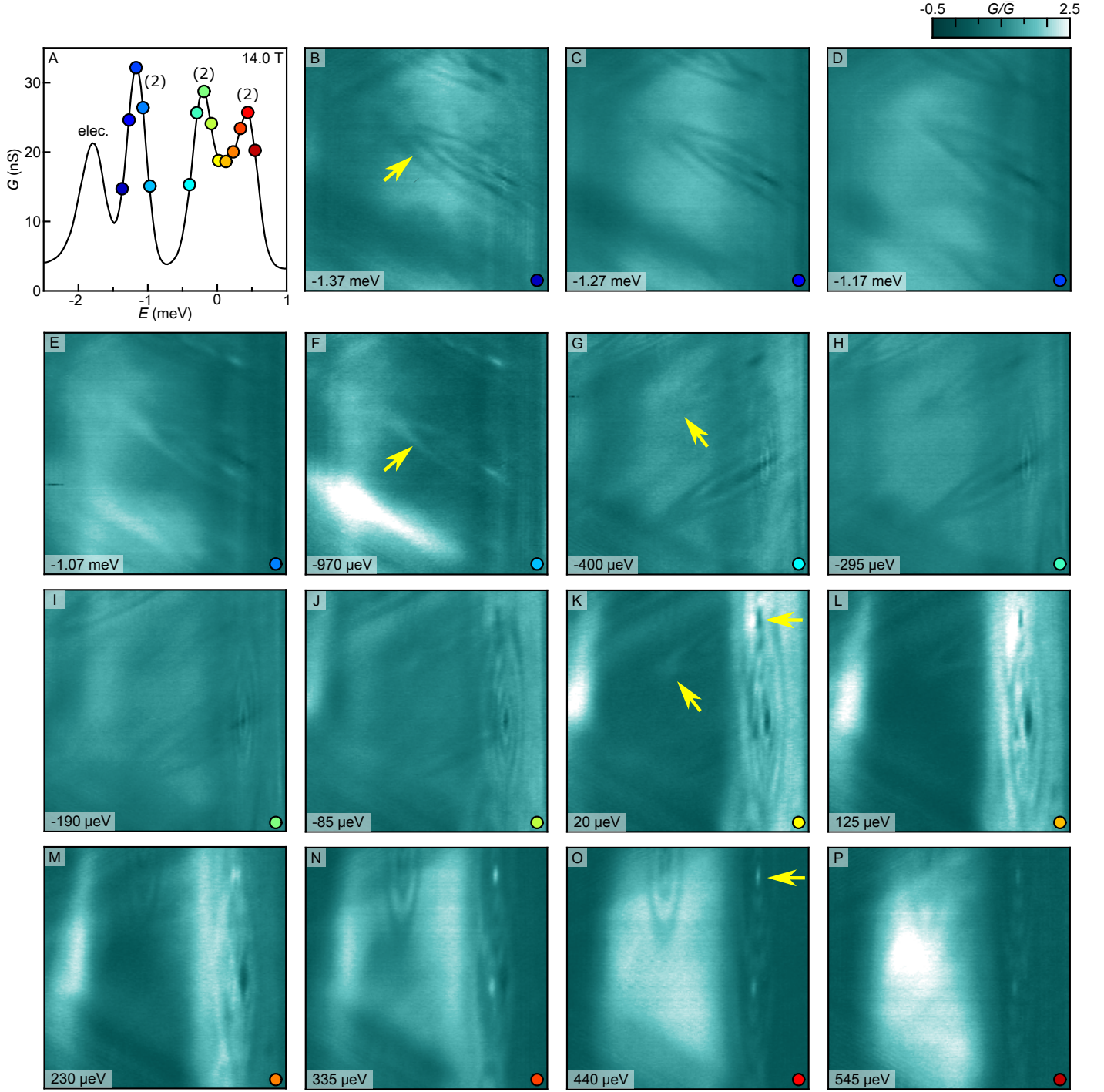

**Fig. S5.** Full spatial and energy dependence of conductance across all three broken-symmetry states in a second location. (**A**) Average conductance spectrum measured at $B = 14$ T in the same location as the data in Figs. 1E and 1G. The $N_h = 3$ LL is split into three peaks, two of which are split by exchange interactions at the Fermi level. The $N_e = 8$ LL is also visible. The degeneracy of each hole LL is labeled in parentheses, and the colored circles mark the mapped energies in panels B-P. (**B**)-(**P**) Spatial maps of $G/\overline{G}$ at $B = 14$ T with energy spaced by approximately 100 µeV throughout the three broken-symmetry $N_h = 3$ LL peaks. In addition to showing the same sequence of nematic orientations as the $N_h = 4$ LL in the same area (Fig. S4), these maps reveal subsurface defects which shift the $m = N$ cyclotron orbit about 420 µeV higher in energy. The energy shift is similar for all three nematic orientations, and the yellow arrows highlight the most prominent ellipses that arise from subsurface defects. For each directionality, an arrow highlights both a signal with suppressed conductance and another (in the panel immediately below) with enhanced conductance relative to the background. We note that the bright ellipse in Fig. S5F is from the subsurface defect, not the nearby raised surface defect (dark ellipses from both can be seen in Fig. S5B).

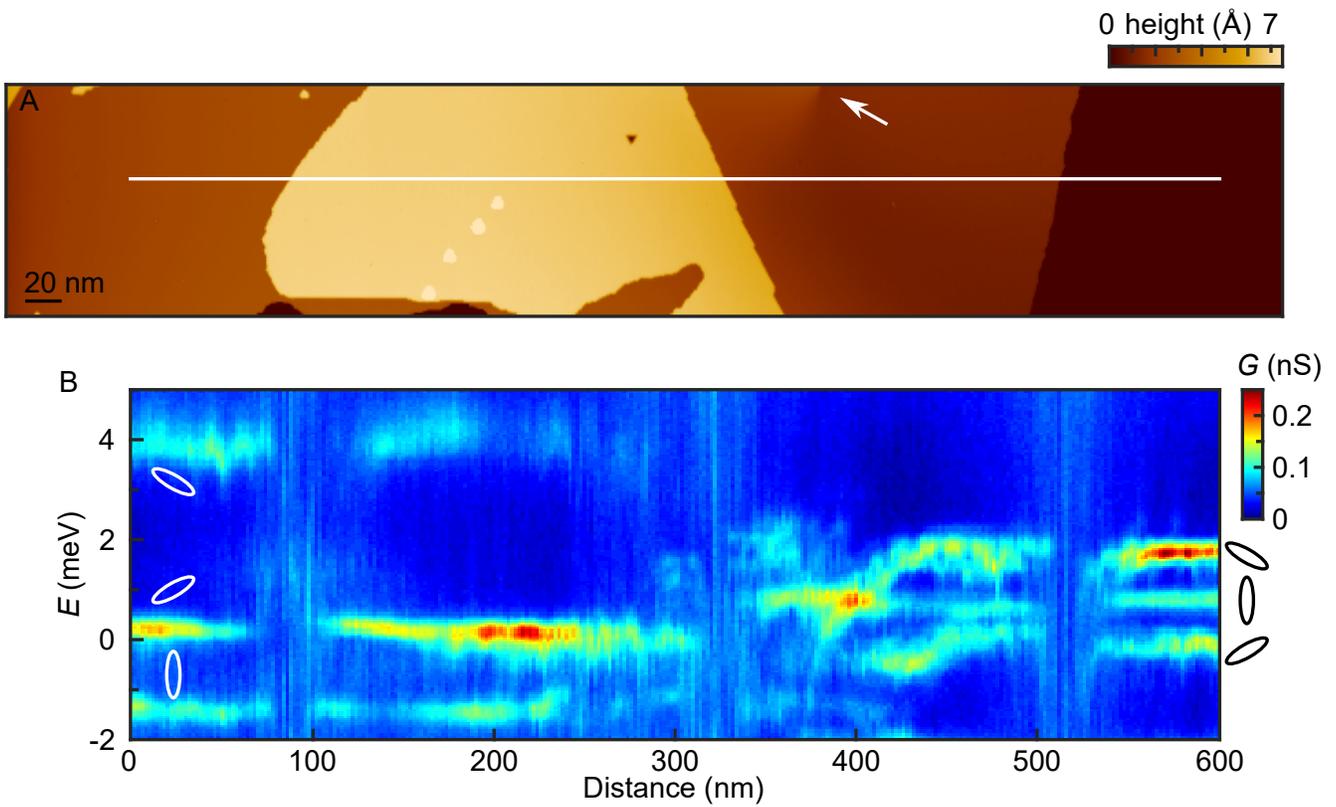

**Fig. S6**. Measurement across a domain wall. (**A**) Topography of the Bi(111) surface. The arrow marks a strain defect on the surface which is visible as a deformation in the topographic signal. The white line shows the trajectory of the spectroscopic linecut, which starts in the domain shown in Figs. 2A-D and ends in the domain shown in Figs. 2H-K. $I_{set}$ = 30 pA. (**B**) Spectroscopic linecut across a domain wall that shows the evolution of the three split $N_h$ = 3 LL peak energies with position. The directionality of the real-space conductance features that we observe for each LL peak is labeled as white and black ellipses, respectively, for the two domains. $V_{rms}$ = 74 µV.

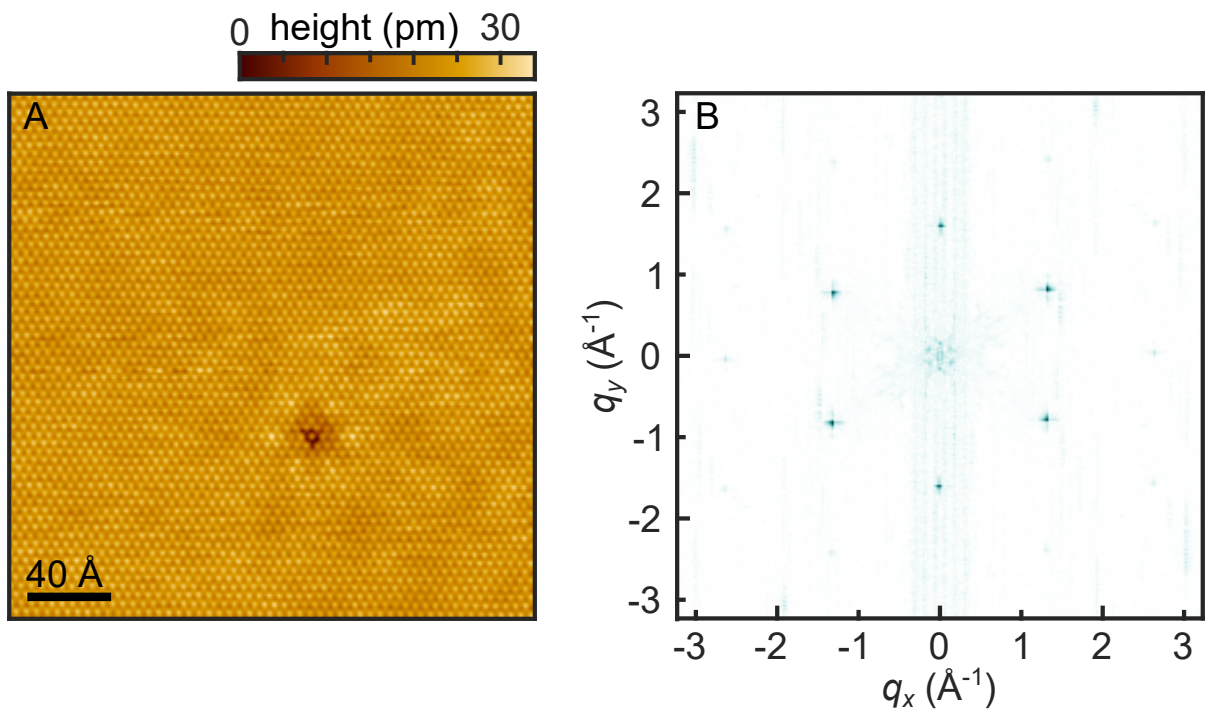

**Fig. S7**. Atomic resolution image of the Bi(111) surface and a defect. (**A**) Topography showing the surface Bi atoms and an isolated defect. The Bi(111) surface is perfectly ordered, and the defect does not break the three-fold symmetry of the lattice. $V_{set}$ = 10 mV and $I_{set}$ = 20 nA. (**B**) Fourier transform of the image in A.

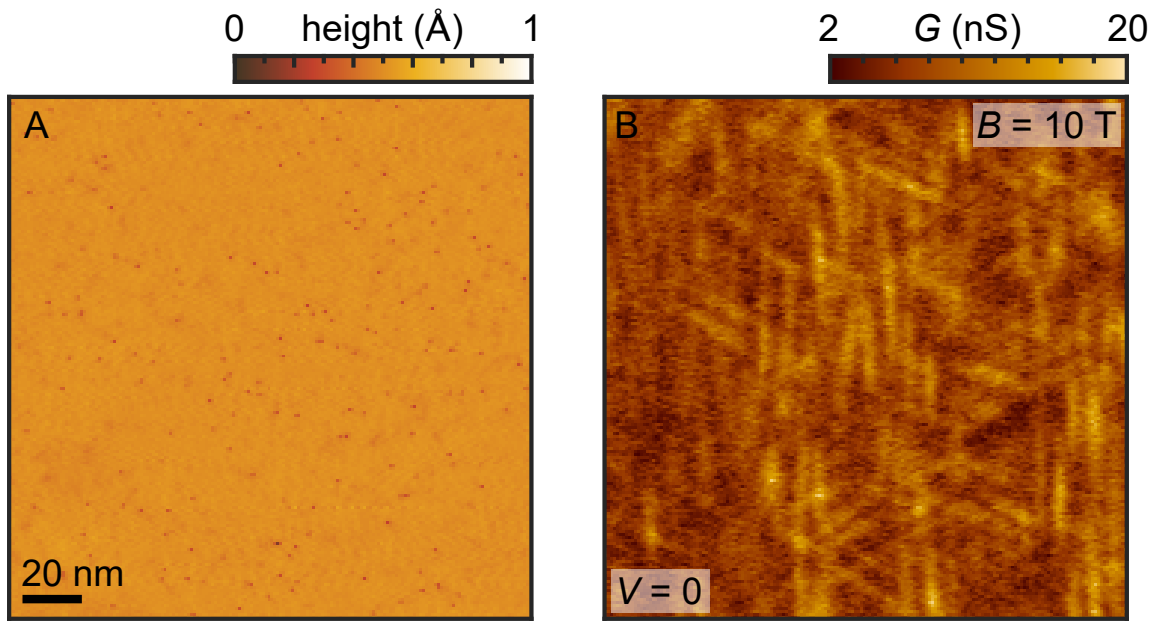

**Fig. S8**. Measurement of a doped Bi sample. (**A**) Topography showing the surface of a doped Bi(111) sample. (**B**) Representative spatial map of $G$ that reveals pinned cyclotron orbits of all three directionalities at a single energy.

**Table S1.** Magnetic field at which hole LLs cross the Fermi level, the corresponding filling factor, and the derived hole density $p$.

| $B$ (T) | $\nu$ | Derived $p$ ($10^{12}$ cm$^{-2}$) |
|---|---|---|
| 14 | 21 ($N_h = 3$ LL half filled) | 7.11 |
| 10.875 | 27 | 7.10 |
| 8.875 | 33 | 7.08 |
| 7.5 | 39 | 7.07 |
| 6.5 | 45 | 7.07 |
| 5.75 | 51 | 7.09 |
| 5.125 | 57 | 7.06 |
| 4.625 | 63 | 7.05 |
| 4.25 | 69 | 7.09 |

**Table S2**. Magnetic field at which electron LLs cross the Fermi level, and three possible assignments of filling factor ($v$, $v_-$, and $v_+$) for each crossing with the corresponding derived electron density. The assignment that best yields a constant electron density $n$ as a function of field is $v$.

| B (T) | v | Derived n ($10^{12}$ cm$^{-2}$) | v- | Corresponding n- ($10^{12}$ cm$^{-2}$) | v+ | Corresponding n+ ($10^{12}$ cm$^{-2}$) |
|---|---|---|---|---|---|---|
| 13.1 | 9.5 | 3.01 | 8.5 | 2.69 | 10.5 | 3.32 |
| 11.75 | 10.5 | 2.98 | 9.5 | 2.70 | 11.5 | 3.27 |
| 10.75 | 11.5 | 2.99 | 10.5 | 2.73 | 12.5 | 3.25 |
| 10.05 | 12.5 | 3.04 | 11.5 | 2.79 | 13.5 | 3.28 |
| 9.2 | 13.5 | 3.00 | 12.5 | 2.78 | 14.5 | 3.22 |
| 8.65 | 14.5 | 3.03 | 13.5 | 2.82 | 15.5 | 3.24 |
| 8.15 | 15.5 | 3.05 | 14.5 | 2.86 | 16.5 | 3.25 |
| 7.6 | 16.5 | 3.03 | 15.5 | 2.85 | 17.5 | 3.22 |